\documentclass[12pt]{iopart}
\usepackage{graphicx,iopams}
\usepackage{pstricks-add}
\usepackage{cite}
\usepackage[normalem]{ulem}
\usepackage[none]{hyphenat}

\def\pmb#1{\setbox0=\hbox{#1}
\kern-.025em\copy0\kern-\wd0 \kern.05em\copy0\kern-\wd0
\kern-.025em\raise.0433em\box0}

\newcommand{\text}[1]{\rm #1}

\begin{document}

\title[Complex systems: physics beyond physics]{Complex systems: physics beyond physics}

\author{Yurij Holovatch$^{1,2}$, Ralph Kenna$^{2,3}$, Stefan Thurner$^{4,5,6,7}$}

\address{$^{1}$
Institute for Condensed Matter Physics, National Academy of
Sciences of Ukraine, UA--79011 Lviv, Ukraine}

\address{$^{2}$
Doctoral College for the Statistical Physics of Complex Systems,
Leipzig-Lorraine-Lviv-Coventry $({\mathbb L}^4)$, Europe}

\address{$^{3}$
Applied Mathematics
Research Centre, Coventry University, Coventry CV1 5FB, United
Kingdom}

\address{$^{4}$
Section for Science of Complex Systems, Medical University of
Vienna, A--1090 Vienna, Austria}

\address{$^{5}$
Santa Fe Institute, Santa Fe, NM 87501, USA}

\address{$^{6}$
IIASA, Schlossplatz 1, A--2361 Laxenburg, Austria}

\address{$^{7}$
Complexity Science Hub, Josefst\"adterstra\ss e  39, A--1080 Vienna, Austria}

\begin{abstract}
Complex systems are characterized by specific time-dependent
interactions among their many constituents. As a consequence they
often manifest rich, non-trivial and unexpected behavior. Examples
arise both in the physical and non-physical world. The study of
complex systems forms a new interdisciplinary research area that
cuts across  physics, biology, ecology, economics, sociology, and
the humanities. In this paper we review the essence of complex
systems from a physicist's point of view, and try to clarify what
makes them conceptually different from systems that are
traditionally studied in physics. Our goal is to demonstrate how the
dynamics of such systems may be conceptualized in quantitative and
predictive terms by extending notions from statistical physics and
how they can often be captured in a framework of co-evolving
multiplex network structures. We mention three areas of
complex-systems science that are currently studied extensively, the
science of cities, dynamics of societies, and the representation of
texts as evolutionary objects. We discuss why these areas form
complex systems in the above sense. We argue that there exists
plenty of new land for physicists to explore and that methodical and
conceptual progress is needed most.
\end{abstract}

\pacs{89.75-k; 89.75.Hc} \submitto{\EJP}

\eads{
\mailto{hol@icmp.lviv.ua},
\mailto{r.kenna@coventry.ac.uk},
\mailto{stefan.thurner@meduniwien.ac.at}
}

\maketitle


\vskip 1cm

\section{Introduction}\label{I}
As a natural science, physics is based on the language of
mathematics, the interplay between theory and empirics, and the
ultimate primacy of experiment. Its foundational purpose is the
predictive understanding and modelling of natural phenomena. Its
prime driver is curiosity. The curiosity of physicists is taking the
discipline to a wider set of fields, tackling problems traditionally
seen to belong to the remits of other sciences and academic domains.
New types of problem areas are being explored. One of these, {\em
the nature of complex systems}, is the subject of the following
thoughts.

The notion of a complex system is gradually emerging to be one of
the inherent concepts of modern science. On a wider scale, it also
appears in sociological and cultural contexts. The continual
expansion of the dominion where the notion of complexity is used, as
well as the discovery or recognition of wider ranges of phenomena
where it is applicable, leads to difficulties in establishing a
strict definition. Sometimes the term is applied to any system
consisting of many interconnected parts which, as a whole, possesses
properties that are not trivial aggregates of the properties of its
separate constituents
\cite{complex_system,Sherrington10}.\footnote{Given the rapidly
evolving nature of the subject of this paper, references to the
literature are intended to connect to some important or interesting
sources, rather than to deliver an exhaustive or complete list.} A
reference point often cited in the context of collective effects of
complex systems is the paper {\em More is different}, written by
Philipp Anderson \cite{Anderson72}, a Nobel Laureate in physics. The
science of complex systems addresses the ways in which the
constituent parts give rise to the collective behaviour of a whole
system. However, such an interpretation has limited usefulness for a
physicist, as it encompasses too broad a set of circumstances.

Historically, another more useful step in defining complex systems
appeared in physics some time ago: a system is complex if its
behaviour crucially depends on its details \cite{Parisi99}. In this
context, one means phenomena such as deterministic chaos, quantum
entanglement, protein folding, spin glasses, etc. Collective complex
behaviour can arise under the influence of frustration and
structural disorder. As a consequence, an equilibrium state is hard
to reach and responses to external perturbations are slow and very
often random \cite{Parisi99,Sherrington10,Goldenfeld99}. Such
very different phenomena are studied in different fields of physics
such as dynamical systems, quantum mechanics and statistical
physics. Their common feature is that infinitesimal changes in
initial conditions (albeit of very different natures) lead to
radically different scenarios in the time evolution of these
systems.

There is a second component that is essential for the definition of
complex systems. On the one hand, interactions between constituent
parts lead to collective behaviour and define the macrostate but on
the other hand, the interactions are modified in the course of the
system's evolution and are influenced by the macrostate. In other
words, the macrostate and microstates dynamically update each other.
Analysis of such effects has lead to the creation of methods and
development of concepts that were successfully applied to the
description of formally similar phenomena occurring in chemical,
biological, social and other systems which are formed by agents of
non-physical origin\cite{Stauffer00,Sornette01}.

Finally, the notion of complex systems needs the possibility that
the interactions between constituents are time varying, and many
different interaction types can be present at the same time.
Interactions between elements can be very specific. These
interactions change the states of the constituent elements. The
essence of many complex systems is that the states of constituents
and interactions co-evolve over time, meaning that interactions
change states of constituents, and the states of the constituents
change the networks of interactions between constituents
\cite{thurnerlecture2010}. If physics is the science of the four
fundamental forces which hold matter together, the science of
complex systems is its generalization to forces and matter of a
broader concept. Forces can be anything that change states of
constituents, matter is anything where a force can apply
\cite{thurnervisions}. Examples include in condensed matter,
ecology, biology, stock markets, economics, sociology and
humanities.

Here we mainly focus on complex systems comprising constituents or
agents which arise beyond physics, and even beyond the physical. We
intentionally omit examples from within traditional physics because
these are adequately covered elsewhere \cite{Mezard87,Stein13}. Our
principal goal is to demonstrate how the behaviours of systems of
agents which originate beyond physics, or even those which have
non-physical origin, may be described in terms of physics concepts
(e.g., universality, scaling, phase transitions, entropy, random
walks, percolation, diffusion, etc.) and how they can be
quantitatively analysed using tools associated with modern
statistical physics. Our paper is primarily addressed to those
sceptical physicists who consider that dealing with systems of
agents of non-physical origin can not be called physics. We believe
that there is a physics beyond the study of planets, atoms,
molecules and energy. We promote the opinion that it is not the
object of study that matters, but rather the system of notions and
methods that are used for its quantitative and predictive (possibly
probabilistic) description \cite{Ball}.

The paper is structured as follows. In section~\ref{II} we discuss
the notion of predictability and how its understanding has changed
in the course of evolution and revolutions in physics. In
section~\ref{III} we formulate what is a complex system and
introduce some observables used for its description. In
section~\ref{IV} we discuss a dynamical-network formalism that
currently serves as a universal language in complex-systems science.
Finally, section~\ref{V}  presents examples of complex systems
outside physics: we discuss evolution of societies, cities, and
culture itself. Conclusions and an outlook are given in
section \ref{VI}.

\section{Predictability in physics}\label{II}

The mode of thinking that has lead to the emergence of a science of
complex systems is closely connected to the way the notions of
determinism and predictability evolved in physics. The concept of
predictability evolved from strict determinism of trajectories to a
probabilistic description in statistical mechanics and quantum
mechanics.

In the early nineteenth century by ``predictability'' of a
many-particle system one meant that the coordinates of all particles
and their states (momenta) should in principle be calculable as
functions of time, no matter how large the number of particles is and
provided that the initial conditions and equations of motion were
known. This understanding is illustrated in the famous statement by
Pierre Simon Laplace in {\em Essai philosophique sur les
probabilit\'es} (see Ref.~\cite{Laplace}, p.4  for an English
translation):
\begin{quote}
``...We ought then to regard the present state of the
universe as the effect of its anterior state and as the cause of the
one which is to follow. Given for one instant an intelligence which
could comprehend all the forces by which nature is animated and the
respective situation of the beings who compose it an intelligence
sufficiently vast to submit these data to analysis, it would embrace
in the same formula the movements of the greatest bodies of the
universe and those of the lightest atom; for it, nothing would be
uncertain and the future, as the past, would be present to its
eyes....".
\end{quote}

Two major changes in understanding predictability of a physical
system lead to (and were caused by) the appearance of statistical
mechanics and quantum mechanics. In the framework of the former,
microscopic states of  many-particle systems are described by
probabilities. There, predictions associated with a certain
observable (including the state of a particle) no longer mean strict
determinism of trajectories, but the determination of quantities
such as mean values, variances, distribution functions, etc. The
latter, quantum mechanics, does not even allow exact measurements of
more than one state anymore. The notion of  ``predictions'', even
for a single-particle state, have now  become of a probabilistic
nature: either they involve mean values of observables for a single
particle from a system (ensemble) of interacting classical particles
or they deal with mean values of observables for a unique quantum
particle measured during a sequence of experiments.

Yet a new change in the notion of predictability (of classical
systems) was caused by the understanding of deterministic chaos,
where it became clear, that for certain classes of dynamical
systems long-time predictions are impossible in practice: small
changes in the initial conditions would lead to enormous changes in
the outcome. Note that equations of motion of such systems do not
contain stochastic terms: they are fully deterministic, however they
are non-linear. For example in weather forecasting, running the same
computer code on different machines can deliver strongly different
outputs due to the accumulation of differing rounding errors during
computations \cite{weather}. These effects are caused by the
non-linearity of equations of motion and can be present even for a
single classical object.

In complex systems we will keep the notion of probabilistic
predictions. What is different to classical statistical mechanics
and quantum mechanics is that the underlying systems not only might
have non-linear interactions, but also interactions are
time-varying. This may lead to much richer phase diagrams and
macroscopic behaviours than we know from statistical mechanics.

\section{What is a complex system for a physicist?}\label{III}

\subsection{Complex vs complicated}\label{IIIa}

Most complicated systems in physics are not complex. Certainly
particle physics is more or less complicated, however it shows no
signs of complexity.  In the standard theory of elementary particles
interactions between particles always happen in the same way.
However, it is true that most complex systems are complicated
\cite{Gell-Mann96}.

Let us consider a collection of classical nodes or entities (e.g.,
spins, agents) connected by links (interactions). Each node can be
in one of a number of possible states described by a scalar, vector,
matrix, or other mathematical object. For example, the state of the
node can be the value of the local magnetic moment of a spin or the
opinion of an agent (god exists or does not exist). The overall
system is also characterized by a state, in a spin system that would
be the overall magnetization, in the social network it could be the
majority opinion. A single configuration of all nodes is a {\em
microstate},  the state of the whole system is the {\em macrostate}.
We first consider a `simple' (not complex) system; e.g., a random
network or a regular lattice. Denoting the state of a node by
$\sigma_i$, one may describe its evolution by a system of equations,
\begin{equation}\label{III.1}
\frac{d\sigma_i}{dt}=F(\{\sigma_i(t)\},{\bf M})\, , \hspace{1em}
i=1,\dots,N\, ,
\end{equation}
where $N$ is the number of agents, ${\bf M}$  is the interaction
matrix with elements $M_{ij}^\alpha$ that span all pairs of agents
and the index $\alpha$ accounts for possible different types of
interactions. The function $F$ can be stochastic or deterministic.
In the latter case, solutions of Eq.~(\ref{III.1}) are uniquely
determined by the interaction matrix and the initial conditions,
$\{ \sigma_1(0), \sigma_2(0), \dots, \sigma_N(0)\}$. In this
sense, the latter may be considered as a `boundary condition' for
the system that determines the differential equations~(\ref{III.1}).
Of course, the equations might be (and usually are) complicated but,
at least in principle, they are analytically tractable. Henceforth,
we may call such systems  `complicated' but not complex. Needless to
say, with an increase in number of agents $N$, methods of
statistical physics will come into play, as briefly outlined in
Section~\ref{II}.

For a complex (as opposed to a complicated) system, the states of
the nodes again determine the state of the system but they also
influence the interactions between the nodes. In other words, the
microstates (nodes) and the nature of the interactions ${\bf M}$
dynamically update each other. The evolution of such a system is
therefore governed by a system of equations,
\begin{eqnarray}\label{III.2}
\frac{d\sigma_i}{dt}&=&F(\{\sigma_i(t)\},{\bf M}(t))\, ,
\hspace{1em} i=1,\dots,N\, , \\ \label{III.3}
\frac{dM^\alpha_{ij}}{dt}&=&G(\{\sigma_i(t)\},{\bf M}(t))\, ,
\hspace{1em} i<j=1,\dots,N\, .
\end{eqnarray}
The principal difference between the systems of differential
equations (\ref{III.1}) and (\ref{III.2}) is that in the former case
the matrix ${\bf M}$ does not change over time (the `boundary
conditions' of the system defined as a set of equations are fixed)
whereas in the latter case evolution of ${\bf M}$ is governed by
Eq.~(\ref{III.3}). Rephrasing the famous chicken-or-egg causality
dilemma, one can ask: do the interactions define the states (as in
the above example of a `complicated' system) or, vice versa, do the
states define the interactions?
It is the second equation (\ref{III.3}) that makes this system
analytically intractable: given the system of differential equations
with the boundary conditions that are not fixed, one is not able to
get an analytic solution. The dynamical update of the two sets of
equations in Eq.~(\ref{III.2}) and Eq.~(\ref{III.3}) is similar to
the dynamics of how an algorithm works. It changes its internal
states (here that would be the interactions) as it evolves. These
processes no longer follow analytic dynamics (solutions to equations
of motion) but {\em algorithmic dynamics}. Physics in its
traditional scheme (from Newton to spin glasses) is a subset of
these equations, namely Eqs.~(\ref{III.2}) alone. In this sense it
becomes very intuitive in what sense complex systems generalize
traditional physics. In some cases the analysis of complex systems
can be (and is) treated by reducing Eqs.~(\ref{III.2}) and
(\ref{III.3}) to a system like Eq. (\ref{III.1}). This can be
achieved for example when the dynamical processes described by Eq.
(\ref{III.2}) and by Eq. (\ref{III.3}) occur on different time
scales.

For example, consider a social system, wherein agents (nodes) are
individuals that interact by means of specific social interactions
(links), like communication, trading, liking, etc. In real life the
`states' $\sigma_i$ in which an individuals  $i$ can be, could
include its wealth, education level, social influence, religion,
etc. For the purpose of describing human behaviour during elections,
it is often enough to consider them being in one of two possible
states: `god exists' and `does not exist' \cite{Galam12}. Clearly,
it is conceivable that my interactions ($\bf M$) with others will
determine or at least influence  who I am going to vote for
$\sigma_i$. At the same time, who I am (typically) voting  for will
have an influence on my future social environment ($\bf M$).
Similarly, when describing the spreading of infections in the
framework of a SIR model, one considers individuals being in one of
three possible states: susceptible to infection, infected, or
recovered with immunity \cite{Kermack27}. Depending on the states
the interactions between individuals are different: if infected
meets recovered the interaction is: pass infection, whereas if two
infected meet, nothing happens, no interaction (and no
state-change). It is not the number of possible states, nor those of
interactions that makes social systems complex, but the co-evolution
of states and interactions.

An example of co-evolution that happens at very distinct time scales
is the transportation system of a city. Consider bus and metro
stations as nodes $i$ and represent the public transportation
network as a set of links between them ($M_{ij}$)
\cite{vonFerber07,vonFerber09}. The transportation network is
changing over time as a function of preferences of individuals using
the system, and new bus lines, stations, bridges are opened, etc.
However, these infrastructural changes evolve much slower than the
typical urban mobility processes.

Spin glasses are frequently cited as examples of physical complex
systems \cite{Mezard87,Sherrington10,Stein13}. Also protein
folding has become an increasingly popular theme in statistical
physics \cite{Parisi99}. For a spin glass, the state of a single
node is described by a (classical) spin variable, $\sigma_i$ which
can be $\pm 1$ in the Ising case. The interactions $M_{ij}$ account
for two main ingredients that lead to the spin-glass state:
frustration and randomness. Within the Edwards-Anderson model
$M_{ij}$ are taken to be random variables with the distribution
\cite{Edwards75},
\begin{equation}\label{III.4}
P(M_{ij})= \sqrt{\frac{1}{2\pi M^2}}\,\, e^{-M^2_{ij}/2M^2}\, ,
\end{equation}
which has zero mean and variance $M^2$. In this representation
frustrations arise through competing ferro- and antiferromagnetic
interactions (note that $M_{ij}$ can have different signs). This
leads to typical features of the spin-glass state: the presence of
many relevant but non-equivalent macrostates, slow reaction to
external perturbations, etc. Although these features are inherent to
complex systems, such a description misses an essential attribute:
the dynamical co-evolutionary update between microstates and
interactions (cf. Eqs. (\ref{III.2}), (\ref{III.3})) is substituted
by a randomness in the interactions $M_{ij}$.
Again, the presence of different time scales allows one to treat
certain types of disordered magnets reducing Eqs.~(\ref{III.2}) and
(\ref{III.3}) to a system of equations of the form (\ref{III.1}). In
particular, this concerns cases when magnetic degrees of freedom in
a structurally-disordered magnet relax much faster than its
structure.\footnote{This last statement concerns only  so-called
quenched magnets \cite{Folk03}. In  annealed magnets both magnetic
and non-magnetic degrees of freedom relax on the same time scale
\cite{Brout59}.}

An example from biology that contains all attributes of the complex
system behaviour is gene expression. There, the fact that a certain
gene is  expressed (meaning that the gene is involved in
the synthesis of a functional gene product) depends on whether some
other genes are expressed.  As soon as gene products are present
a series of specific non-linear interactions may take place that
change the expression levels of yet other genes. In other words, the
state of a gene (its expression) is dynamically updated with the
interactions that regulate the states of other genes
\cite{Barabasi04,Dehmer11}.

A truly complex physical ``system'' in the above sense is  general
relativity. There, the already mentioned chicken-and-egg dilemma is
obviously present: the dynamics of the agents (their trajectories in
space) unfolds as given by Eq. (\ref{III.2}) for a fixed space  $\bf
M$, however as a result of the dynamics, the space $\bf M$ changes
itself according to  Eq. (\ref{III.3}). We will consider three
examples of this ``algorithmic''  type in detail in the following
section, however outside traditional realm of physics. Before
doing so we introduce the notions of power law statistics and
measures of complexity.

\subsection{Power laws}\label{IIIb}

Complex systems are almost never characterized by Gaussian
statistics. This is due to the fact that the central limit theorem
often no longer holds for systems that are strongly interacting,
path-dependent and non-ergodic, as those that are governed by our
Eqs. (\ref{III.2})-(\ref{III.3}) typically are. Instead, the
statistics of complex systems are dominated by fat-tailed
distributions and very often these are power laws. Depending on
context and exponents, these are known under various names, such as
student t-distributions, Zipf, Pareto, Mandelbrot, Cauchy, Lorenz,
or Tsallis distributions, just to mention a few.

Often the distribution functions that are available are not
probability distribution functions of measured variables (such as
the velocity distribution of gas particles) but are frequency
distributions. A famous example is the analysis of word frequencies
$f$ in texts. When all words of the text are ordered by descending
frequency one finds  \cite{Zipf35}\footnote{George Zipf was not the
first who noticed the power-law decay in the function (\ref{III.5}).
Earlier observations were made by J.~B.~Estoup (in 1916) and by
E.~U.~Condon (in 1928) \cite{Mitzenmacher03,Newman05,Simkin11}.},
\begin{equation} \label{III.5}
 f(r)=A r^\kappa.
\end{equation}
where  $A$ is a normalization constant and the power exponent
$\kappa$ was for a long time considered to be the same and universal
for all languages, $\kappa\simeq -1$  and independent of factors
such as the author, genre, time when the text was written, etc. This
is no-longer believed, see \cite{Thurner15} and the references
therein.

Power laws that govern statistics of systems of many interacting
agents were discovered in different disciplines at different times.
Very often these laws hold names of their discoverers: in economics
it is the distribution of wealth among individuals (V. Pareto,
1896); in demography  -- distribution of towns according to their
size (F. Auerbach, 1913); in biology -- distribution of sizes of
biological genera according to the number of species they contain
(J.C. Willis, G. Yule, 1922);  in scientometrics -- distribution of
papers written by separate scientists (A.J. Lotka, 1926), of
scientific journals according to the number of papers they contain
(S.C. Bradford, 1934), citations (D. de S. Price, 1965). The list is
terribly incomplete and goes on.

While in most cases where Gaussian or Boltzmann distributions occur,
the Central Limit Theorem is at work in one way or the other, the
origin of power law distributions is not so simple. The wide range
of phenomena that exhibit fat tailed distributions suggests that the
reasons for their appearance have to be quite general and should not
depend on the individual peculiarities of their constituting parts
or interactions. On the other hand, the fact that different
phenomena are described by the same power law does not mean that
the origin of these laws is the same. The origin of power laws can
be explained by several very different mechanisms. In particular, the
main routes to power laws are
\begin{itemize}
\item critical phenomena \cite{Kadanoff67,Fisher98,Itzykson89},
\item preferential processes \cite{Yule25,Simon55,Barabasi99},
\item self-organized criticality \cite{Bak87},
\item multiplicative processes with constraints \cite{Mitzenmacher03,Newman05},
\item optimisation  \cite{Mandelbrot53},
\item path dependent, non-ergodic processes that reduce their sample space as they unfold \cite{Corominas-Murtra15}.
\end{itemize}
A typical example of the appearance of power laws in condensed
matter physics is given by critical phenomena. As the critical point
is approached, universal power laws govern  divergencies in several
thermodynamic observables (like compressibility of fluids or
susceptibility of magnets) as well as structural (large distance
asymptotics of the pair correlation function) and statistical
(correlated cluster size distribution) properties
\cite{Kadanoff67,Fisher98,Itzykson89}. There, an increase of
fluctuations in the fine-tuned vicinity of the critical point
provides a mechanism for the emergence of power laws.

Preferential processes that are sometimes referred to as the ``the
rich get richer'' phenomenon were suggested by Yule and later by
Herbert Simon \cite{Simon55}. If outcomes of processes occur
proportional to the number of times they have occurred in the past,
power laws in the occurrence frequencies appear. This mechanism has
gained recent popularity in the form of the preferential attachment
mechanism in the formation of scale-free networks \cite{Barabasi99}.
With preferential attachment, in the course of system evolution, new
elements tend to create links with those, who already have more
links.

Another class is self-organized criticality, where power laws appear
due to non-linear interactions between system constituents
\cite{Bak87}. The system adjusts automatically to the critical point
and no fine tuning is necessary. The classical example is the slope
that self-organizes in a sandpile when sand is dropped grain by
grain on a surface.

Mutliplicative processes (products of sequences of random numbers)
lead to lognormal distributions, that sometimes are hard to
distinguish from true power laws in data. This is a trivial result
of the Central Limit Theorem when applied to logarithmic variables.
If multiplicative processes are subjected to simple constraints they
can very easily show true power laws, as reviewed for example in
\cite{Mitzenmacher03}.

An optimization scenario that leads to power laws was suggested by
Benoit Mandelbrot, which is based on information theoretical
arguments. In this scenario, power laws appear as a result of the
optimization of the costs associated with the transmission of
information.

Recently it was understood that all history-dependent processes that
become more constrained as they unfold (sample-space reducing
processes) lead to power laws. It constitutes an independent class
of processes that lead to the power law statistics
\cite{Corominas-Murtra15,Thurner15}.

All the above mechanisms are present in our framework of   Eqs.
(\ref{III.2}) and (\ref{III.3}) as limiting cases. It remains a
challenge to see which of these processes is dominating for a given
complex system at hand. Very often for a specific complex system it
might be a mixture of these processes that lead to power laws.

\subsection{Measures of complexity}\label{IIIc}

As the notion of complexity  entered the domain of the mathematical
sciences, there were attempts to define quantitative ``measures of
complexity''. The current abundance of such measures is caused
partly by the fact that they were introduced in different fields and
that they quantify different aspects of complex systems. Many of
these measures are similar and show strong overlaps, so that one can
try to  group these measures into different taxonomies. Currently,
there exist different classifications of complexity measures,
depending on which features are chosen to be essential for a given
group. One classification is suggested by Lloyd \cite{Lloyd01} where
measures are grouped according to the questions they are supposed to
answer, namely
\begin{itemize}
\item (i) how hard is it to describe?
\item (ii) how hard its is to create?
\item (iii) what is its degree of organization?
\end{itemize}

The degree of difficulty in describing a complex system completely
(i) is usually quantified with measures such as  information,
entropy and algorithmic complexity  that is sometimes called
Kolmogorov complexity \cite{Kolmogorov65,Kolmogorov83}. Measures to
answer how hard it is to create a complex system (ii) include
computational complexity, logical \cite{Bennett88} and thermodynamic
\cite{Lloyd88} depth, and cost. The answer to the degree of
organization in a system  (iii) is quantified in measures like
effective complexity \cite{Gell-Mann96}, fractal dimension, and
stochastic complexity \cite{Rissanen89}. Other measures of
complexity that are sometimes used are classified as: (i)
non-computable vs. computable and (ii) deterministic vs statistical;
see e.g. \cite{Ladyman13} for more details.

In our opinion several new measures are needed that capture the
degree of {\em co-evolutionariness}. If it is low, traditional
physics can be used. If there exists separation of characteristic
time scales at which the dynamics in Eqs. (\ref{III.2}),
(\ref{III.3}) unfolds, i.e. time scales that can be clearly
separated, co-evolutionariness is low. If it becomes hard to
disentangle dynamics of $\sigma$ and ${\bf M}$ in Eqs.
(\ref{III.2}), (\ref{III.3}) co-evolutionariness is high. This is
the challenge and promises new land for physicists. In terms of
conceptual difficulty it is not more complicated that general
relativity.

\section{Dynamical multilayer networks}\label{IV}

In mathematics a network is a graph comprising a set of vertices and
a set of edges. In physics one often uses the terms nodes and links
instead \cite{Albert02,Dorogovtsev03,Newman06}. In the science of
complex systems networks play a central role because they offer a
way to describe different types of interactions specifically between
agents (not everyone interacts with everyone else). Interactions
that change over time, stochastic interactions, interactions that
occur on multiple levels and that are not embedded in Euclidean
space can all be described within a network formalism.

Depending on the type of interaction, a network can take the form of
an undirected or directed graph. In the former case, interactions
are bi-directional, as is the case for physical interactions, or for
example include scientific-collaboration networks, where two
scientists are linked if they co-author a paper. Directed networks,
on the other hand, arise when relationships between nodes are not
symmetric or mutual  \cite{Dunne02}. For example,  food webs that
describe which agents (species) eat other agents (other species) are
directed, people eat chicken, but chicken don't eat humans. The
strength of interactions can be indicated by the weight of a link in
a graph. One considers unweighted and weighted graphs. For example,
in a transportation network, connectivity simply marks the presence
of a link (e.g., a road) and can be represented by an unweighted
network. Traffic capacities or loads may vary from road to road and
links may be characterized by different weights \cite{Galotti15}.

To describe the situation where many different types of interactions
are simultaneously present between constituents or nodes, one may
use the notion of multilayer networks
\cite{Szell10a,Szell10b,Boccaletti14,Bianconi15,multilayerplex}.
Sometimes a distinction is made between multilayer networks and
multiplex  networks or `networks of networks'. In so-called
multiplex  networks, the same set of nodes is connected by links of
different types. In  `network of networks' nodes of different
networks are connected by inter-links. For example, a connected set
of energy-supply sources (e.g., electric power stations) and another
connected set of computers that control them, together form an
example  of a network of networks. In this case, inter-links between
the two sets determine not only the transmission of control signals
but also the power supply \cite{Rosato08, Buldyrev10}. The relevance
of the topology of the resulting network of networks becomes clear
if one analyzes the reactions of the entire system to the random
removal of some of its nodes. One finds that undesired collective
cascading dynamics can lead to propagation of a failure and
dysfunction of the network as a whole \cite{Buldyrev10}. The
cascading is due to feedback between both interacting networks; the
communication network of computers controls the network of power
stations, which in turn control energy supply to the computers. This
is an example of a percolation phenomenon on interconnected networks
with the uncommon features of an abrupt first order phase transition
\cite{Parshani10,Herrmann16}\footnote{Usually percolation phenomena
have  a continuous phase transition \cite{Stauffer91}.}.

Prior to the recent emergence of network science, a mathematical
theory had been developed for the simplest of networked objects,
so-called Erd\"os-R\'enyi, or  random graphs \cite{Bollobas85}.
These are characterised by a degree distribution which is Poisson,
for large graphs. The {\emph{degree}} $k$ of a node is the number of
links attached to it and the distribution of degrees  across nodes
is one of the most fundamental characteristics of the network
structure. Poissonian degree distributions are not observed in many
real-world networks. Instead typically fat tailed degree
distributions are observed, sometimes they are real power laws,
\begin{equation}\label{IV.1}
P(k) \sim k^{-\lambda} \quad ,  k \gg 1,
 \label{fatty}
\end{equation}
where $P(k)$ is the probability that a randomly picked node in the
network has exactly degree $k$. Typically $k \sim 1-3$. Networks
that are characterized by a power-law decay of the node degree
distribution are called {\em scale-free}; sometimes they are also
called ``complex networks''. This has of course nothing to do with
the  complexity (or simplicity) of the underlying system.

Scale-free networks are often characterised by very short average
distances between randomly chosen pairs of nodes, a circumstance
sometimes referred to as a {\em small world}. Networks with
scale-free structures that underly complex systems may have a strong
impact on the dynamics of the system and may effect percolation
properties or self-organization. Note that complex networks by
themselves are not necessarily ``complex systems''.

In physics we frequently consider systems that are homogeneous and
isotropic. If interactions take place on networks the resulting
``physics''  plays out in a very different environment. The
fat-tail behaviour of degree distributions of interactions
(\ref{IV.1}) reflects strong inhomogeneities. Nodes with a high
degree (hubs) can play very different roles in a network than those
with a low degree. Another important observable that is used to
capture the inhomogeneities in interaction networks is the the
so-called {\em clustering coefficient} $c_i$ of node $i$, which
is  is defined as the proportion of neighbours of $i$ that are
mutually linked. A power law decrease of the clustering $c_i$ as a
function of the degree $k_i$, is a signal for a
{\em hierarchical} organization in the network \cite{Ravasz}. The
degree of clustering has severe dynamical consequences for complex
systems that are built on such networks.

Another useful quantity is the {\emph{assortativity}} of the
network. This is the extent to which similar nodes are mutually
linked. It is measured for example by using Pearson's correlation
coefficient. The {\em degree assortativity} $r_k$ of a network
measures the correlations between the degrees of those nodes which
are connected by a link \cite{Newman2002}. If  $r_k >0$ the network
is called assortative and if $ r_k < 0$ it is said to be
disassortative. Assortativity measures the extent to which similar
vertices associate with each other and, again, has impact on how
interactions in a complex system manifest themselves and evolve over
time.

\section{Examples of physics outside physics and the physical}\label{V}

\subsection{Societies as complex systems -- a new generation of computational sociology}
Societies are obvious complex systems in the sense of
Eqs.~(\ref{III.2}) and (\ref{III.3}). They are composed of many
individuals and institutions $i$ that can be seen as nodes in a
multiplex network. Nodes are described by state vectors that can
represent wealth, education, social roles etc. The set of all states
of a node can be collected into a state vector $\vec \sigma_i$.
These state vectors are not static but change over time, depending
on the interactions the nodes are involved in. In societies, these
interactions $M_{ij}^{\alpha}$ that always happen between
individuals (or institutions) $i$ and $j$,   can be of very
different types, $\alpha$. For example, interaction type $\alpha=1$
could represent communications, $\alpha=2$ trade, $\alpha=3$
friendship, $\alpha=4$ family relation, $\alpha=5$ collaboration,
etc. Interactions are not static but evolve over time and often
explicitly depend on the states of the nodes. Often interactions are
discrete and happen in interaction events.

The complexity now arises through the coevolution  of interactions
and state changes. If one considers the act of trading for example,
typically a trade interaction is is preceded by a sequence of
communication events (deciding on a price for a particular good).
Once the interactions of the trade happens the states of the
involved parties change, the state of wealth in terms of cash of the
buyer diminishes, while the state of owning a particular quantity of
the traded good increases for the buyer. Trading interactions change
states of cash positions, communication events change the state of
informedness of communicators, hostile interactions may result in
changes of the state of health,  etc. Note that interactions do not
happen all the time and are often very specific, and depend on the
states of the involved nodes. No trading event will happen if the
state of wealth of a potential buyer is below the minimum price
expected of the seller.

Interactions do not happen independently of each other; they
mutually influence each other. The networks of one  interaction type
$\alpha$ influences the link formation and destruction in an other
layer $\beta$. In this sense  there exists a network-network
interaction, a completely unexplored field that could open new and
wide territory for physicists.

This notion of complexity in the dynamics of societies is of course
useless in a scientific sense without the ability to measure and
track state changes and interaction events. With the availability of
new generation of data this seems indeed to be possible. For example,
since several decades every trade in financial markets is recorded.
Data on every single payed medical intervention  between patients
and doctors within a country is becoming available
\cite{thurnerPNAS}, and practically every social interaction and
state change amongst human beings can be recorded in the framework
of massive multiplayer online games. In such games where players, in
the form of avatars, live an alternative life in a virtual universe.
Life there is open ended  and extremely high dimensional. Avatars
pursue economic activities (such as being employed or being
entepreneurs), they buy goods and services from other players, and
they form social relationships on various levels such as
friendships, clubs, gangs, parties, kingdoms, countries, financial
systems, etc. All interactions between all players and all their
state changes in wealth, skills, education, leadership role, etc.
can be monitored and recorded completely. In the case of the game
Pardus.at these data have been used to study a society of about
500.000 individuals, as a complex systems in the form of
Eqs.~(\ref{III.2}) and (\ref{III.3}) \cite{Szell10a}. The amazing
situation arises that every component of these equations is recorded
at all times.  The state vectors $\vec \sigma_i(t)$  as well as all
the interactions $M_{ij}^{\alpha}(t)$ are available, where $\alpha$
stands for the different interaction types such as friendships,
trades, enmity relations, communications, hostilities, revenge, etc.
In the particular case of the Pardus.at game, time $t$ ranges over
more than ten years, with a time resolution of a second.

These kind of massive data brings the science of complex systems to
an entirely new quality level. By looking at how Eqs.~(\ref{III.2})
and (\ref{III.3}) unfold over time it becomes possible to develop a
quantitative feeling of how societies work. In particular, the
co-evolution aspects of interactions and state changes can for the
first time be visualized and in a next step be understood. It is
needless to say that this offers an opportunity to understand human
behaviour and social dynamics on a level  of precision  that was so
far only reached in the natural sciences. In the particular case of
the Pardus.at game first steps were taken in this direction in
\cite{Szell10a,Szell10b,pardus1,pardus2,pardus3,pardus4,pardus5,pardus6,Thurner12,Mryglod15,pardus7}.
This situation tempts one to state that for the first time it
becomes possible to turn sociology into a quantitative and
predictive science, a statement, that most sociologists would -- of
course --  object to. For physicists, however, there seems to be new
land ahead that needs conceptual as well as methodological progress.

\subsection{Cities as complex systems}

Another archetypal example of a complex system arises when one
attempts to apply quantitative analysis to understand inherent
features of a city as a whole, its organization, development, impact
and correlations between global processes that define city life. In
fact, an analogy between behaviour of a living system and a city has
far reaching consequences. One of the first arguments in favour of
such analogy can be found in Aristotle `Politics' \cite{Aristotle}.

In the context of what has  been said above, interpretation of a
city as a complex system involves considerations of nodes of various
types (e.g. subway stations in the public transportation network,
houses linked into the energy production and consumption network,
gas storage facilities etc.), forming inter-connected and
co-evolving networks (streets, information systems, traffic routes,
distribution systems such as energy and waste infrastructures, and
other infrastructure facilities). Much in cities has to do with
distribution of people, goods and services, similar to the way in
which nutrients are distributed through the body to reach every cell of an
organism or, indeed, an ecosystem. The co-evolutionary aspect of
city dynamics as described by Eqs. (\ref{III.2})--(\ref{III.3}) is
obvious. Success or failure of distribution events may change the
states of the various nodes. In turn, changes of states in the nodes
may lead to changes in the distribution networks. For example,
chronic overloading of a subway station might eventually lead to the
construction of a new subway line. On shorter timescales the
information provided by a navigation device on traffic density in a
part of a city might lead to changes of route planning for
individual drivers. This may lead to congestion in other parts of
the system: the traffic-flow network on a city's street network is
changing as a consequence of an information flow of current traffic.
These examples, and many more, obviously follow a co-evolution
dynamics as described in Eqs. (\ref{III.2})--(\ref{III.3}).

If one considers the growth of a city as its expansion in
geographical space, this poses a challenge to quantitative
approaches. City growth has been shown to exhibit self-similar (or
fractal) patterns, an observation that might imply a universality of
processes that drive city agglomeration and clustering
\cite{Batty94,Batty08,Batty12}. Several physical growth processes
that are known to lead to such geometry (percolation or diffusion
limited aggregation) have been exploited to explain such growth in
cities \cite{Makse95,Batty08,Batty94}. Fat-tailed distributions that
very often govern statistics of complex systems make `rare events'
to be essential for emerging system properties. In particular,
inhomogeneities in structure lead to another typical feature of
complex systems: their resilience to random failures and their
vulnerability to targeted attacks. In the context of public
transportation this issue was studied in various cities
\cite{vonFerber09,Berche09,Berche12a,Berche12b}.

Obviously, urban growth is much more than expansion in space. Many
observables that  may be used to quantify various aspects of the
evolution of a city are governed by power laws. Some of the observed
exponents are listed in table \ref{tab1}. Power laws in various
urban indicators as a function of city size (i.e. number of
inhabitants) are ubiquitous
\cite{Bettencourt07,Bettencourt10,Bettencourt11,Arbesman09,Schlaepfer14,Batty08},
signalling that complex economic, demographic and social processes
take place in cities. An understanding of the origin of these power
laws is only partially complete. Data about scaling of various city
indicators are universal in the sense that similar exponents are
observed for cities of very different historical, cultural and
economical backgrounds. Another interesting observation that comes
from table \ref{tab1} is that the data seems to  fall into two
``universality classes''
\cite{Bettencourt07,Bettencourt10,Bettencourt11}: (i) indicators
that describe social interaction of some kind (upper part of the
table) manifest super-linear scaling with exponents $\beta \sim 1.15
> 1$, and  (ii) indicators that describe the infrastructure which
typically scale sub-linearly, $\beta \sim 0.85 <1$.

Returning to the analogy of cities and living organisms, it is worth
noting that  numerous physiological characteristics of different
organisms scale with body mass as power laws with exponents that are
multiples of 1/4. A theoretical explanation for this particular
linear scaling laws has been provided \cite{West97,West99}. An
essential ingredient of the theory is the evidence of hierarchical
branching networks that terminate in size-invariant units. It is the
structure of these networks that leads to the universal allometric
scaling observed in nature. Similarly, scaling laws observed in
various city indicators might turn out as  manifestations of city
network structures, and of universal patterns of human social
interaction networks.

\begin{table}[b]
\centerline{\begin{tabular}{l c} \hline
$Y$ & $\beta$ \\
\hline
Number of new patents    & 1.27         \\
Number of inventors   & 1.25         \\
Private R\&D  employment   & 1.34          \\
``Supercreative'' employment   & 1.15         \\
Number of R\&D establishments   & 1.19         \\
R\&D employment   & 1.26         \\
New AIDS cases   & 1.23         \\
Serious crimes   & 1.16         \\
Number of Gasoline stations   & 0.77         \\
Gasoline sales   & 0.79         \\
Length of all electrical cables   & 0.87         \\
Road surface of city  & 0.83         \\
\end{tabular}}
\caption{Scaling exponents for several urban indicators $Y$ vs. city size
$N$: $Y \sim N^\beta$, where $N$ is  population size. See \cite{Bettencourt07} for the data
sources. \label{tab1}}
\end{table}

Big efforts are being  currently undertaken to analyze networks and
timeseries of city data. Examples include analyses  of creative
output of a city (measured e.g. in innovations found in a city
\cite{Arbesman09}), of mobile phone records
\cite{Schlaepfer14,Palchykov14,Martinez-Cesena15} or taxi movements
\cite{Sagarra15} as proxies for spatio-temporal distributions of
people, etc. In fact, cities are data-producing entities generating
data on the level of nodes or agents of different type and the level
of links that are formed through their interactions. Co-evolutionary
aspects of changes in states of agents and simultaneous changes in
their interaction network structure (as described by Eqs.
(\ref{III.2}), (\ref{III.3})) are not yet systematically studied. A
simple example are traffic guiding systems where traffic lights that
govern the traffic, are themselves governed by local traffic loads
\cite{Helbling08}.

The science of cities in terms of dynamical and co-evolving complex
systems as we define them here has only begun. It is largely
scientific new territory. The availability of increasingly complete
datasets will make it possible to make this science
``experimental'', in the sense that it can be formulated in
quantitative, predictive and eventually testable ways.

\subsection{Complex systems in humanities}

History itself may be considered to comprise a complex system. One
can consider the nodes of such a system to be individuals that
existed, still exist or even those that might have existed. They can
again be characterized by states which can, for example, be social
functions in a society, such as being a king, a revolutionary, a
writer of political texts, a demagogue, a simple voter, a
public-opinion former (e.g., newspaper journalist) or similar. They
are linked to each other through social ties like family relations,
marriage, friendship, business and so on, or even through conflict
or competition. Forms of interactions may then be captured in a
multiplex network and layers include kinship, political coalitions,
financial dependencies, taxation, information flow networks through
letters, emails or even through fighting.

Networks were of importance to thought and culture long before the
invention of modern telecommunications systems and examples include
in ancient Icelandic and Scandinavian literature, through Uppsala
romanticism, the Flemish movement, and French modernists, to more
modern Norwegian poetics \cite{Kramarz-Bein}. Efforts have been made
by medievalists to map communication networks over time in this
context \cite{preiser-kapeller}. The notion of place has been
explored in literature both as geographic and narrative phenomena
\cite{Kinniburgh}. Another perspective is to represent historical
records in the form of books or annals into co-evolving structures.

In many texts such as novels, chonicles, sagas, etc, one can
identify individuals  together with their social ties, dependencies
and influence with each other. That it is possible to extract
meaningful social networks from such texts was demonstrated, for
example, in the context of ancient myths in
Refs.\cite{MacCarron12,MacCarron13}. There it was shown that epic
narratives from the world of mythology have certain universal
characteristics. The exploration of such characteristics was
inspired by the notion of universality from the study of critical
phenomena in statistical physics. For example, the networks tend to
be highly clustered, structurally balanced, hierarchical small
worlds with right skewed degree distributions and coherent community
structures. These are properties very similar to the social networks
that we find ourselves in today.

Does that mean that such ancient stories are based on reality? We
don't know for sure, however we can make sensible investigations
by considering such systems as physicists. For example, it
was found that although the famous Anglo-Saxon epic {\em Beowulf}
has many properties of real social networks, it lacks one crucial
one: assortativity. This is the tendency for similar nodes to
connect to each other. In our society, popular people tend to know
each other and the few friends of less popular individuals tend also
to have few acquaintances. So what would it take to make
{\emph{Beowulf}} assortative? In other words, how far is the
narrative away from appearing realistic?

To investigate, a defining characteristic of complex systems was
exploited. As discussed above, for such systems, changes of states
of individual nodes may lead to macroscopic changes in the entire
network. In Refs.\cite{MacCarron12}, the eponymous protagonist
Beowulf was removed from the social network of the narrative
{\em Beowulf}, and the properties of the remainder were
determined. The resulting network was (marginally) assortative,
meaning that it had all the properties of a real social network.
Intriguingly, although the narrative is embellished with obvious
fiction, archaeological excavations in Denmark and Sweden support
the historicity of some of the characters in {\em Beowulf}. The
lead character Beowulf is mostly not believed to have  existed,
however \cite{Klaeber, Chambers}. Thus it is entirely possible that
the character Beowulf is himself fictional but that the backdrop to
the narrative -- the society within which the story unfolds -- is
realistic. It is interesting, and encouraging, that the network
analysis which is based on considering the society as a complex
system, delivers the same conclusion as coming from traditional
humanities considerations \cite{Klaeber, Chambers}.

With the availability of electronic libraries, it is possible to
perform similar explorations to practically any text that involves
people or institutions. Not only can these texts then be studied in
terms of their co-evolutionary dynamics of `states' of individuals
and their interactions, but also individuals that were identified in
several distinct texts can be brought into relation through a
multiplex network. A classic example is the Icelandic sagas. These
contain an abundance of characters, some of whom appear in more than
one narrative. The main characters in one often  appear as minor
ones in another. In text $\alpha$ the local social network of a
given character represented by node $i$ is given by the matrix
elements $M_{ij}^{\alpha}$. We can then combine societies to obtain
an enormous network and to consider its properties. By  seeking to
compare communities in the combined network with individual sagas,
one can speculate as to whether one was a source for another
\cite{MacCarron13}. Similarly, the consistency of social networks
across texts could, for example, be useful to speculate to what
extent a given text was embellished; the amount of inconsistencies
across layers would offer a new approach to historical research.

There is no reason why such modes of thinking could not be extended
to every character that ever appeared in a text. This would offer an
entirely new way in understanding the written history about
humankind.

\section{Conclusions and outlook}\label{VI}

If physics is the science of matter and its interactions, the
science of complex systems is its natural extension where both
matter and interactions are seen in a much broader context.
Interactions can be anything that leads to a change of a state in a
constituent in a complex system, and matter is anything that can
have at least two states and is able to interact. In this sense
complex systems are a natural extension of physics. The framework
that complex systems are co-evolving multiplex networks
\cite{thurnerlecture2010}, where interactions between elements
change states of these elements and where the collection of states
in a system changes the interaction networks, is similar in spirit
as the framework of general relativity, where dynamics needs a space
that is itself changed as a function of the dynamics. Needless to
say that the dynamics of such systems will be impossible to be
solved analytically and without use of massive computational power.

The availability of more and more data in disciplines and fields
beyond physics allows us to observe the states (and their changes)
of elements and their interactions (and their changes) in great
detail. With the electronic fingerprints we leave practically
everywhere, we are getting toward a situation where we have  -- more
and more often -- even complete information on dynamical complex
systems. For many systems we already have such information in the
sense that literally all state changes and all interaction events
are recorded. This situation will make it possible in the next years
to transform the science of complex systems, from presently a
collection of computational and network based methods, into a fully
experimental science.

\vspace{1cm}

We thank Karoline Wiesner for useful discussions.  This work was
supported in part by FP7 EU IRSES projects No. $612707$ ``Dynamics
of and in Complex Systems", No. $612669$ ``Structure and Evolution
of Complex Systems with Applications in Physics and Life Sciences",
and by the COST Action TD1210 ``Analyzing the dynamics of
information and knowledge landscapes".  ST acknowledges
feedback from students of the course: Introduction to Complex Systems,
where many of the co-evolutionary concepts were originally presented.

\section*{References.}

\providecommand{\newblock}{}

\end{document}